\documentclass[
reprint,
superscriptaddress, 
amsmath,amssymb,
aps,
]{revtex4-2}
\usepackage{comment}
\usepackage{graphicx}
\usepackage{dcolumn}
\usepackage{bm}
\usepackage{multirow}
\usepackage{hyperref}
\hypersetup{
    colorlinks=true,
    linkcolor=blue,
    filecolor=magenta,      
    urlcolor=black,
    citecolor=blue, 
    pdftitle={DIS2023 Proceedings},
}

\usepackage[mathlines,modulo]{lineno}
\usepackage{booktabs}
\usepackage{sidecap}
\newcommand{\dd}{\mathop{}\!\mathrm{d}}

\begin{document}
\preprint{APS/123-QED}

\title{Complete Formalism of Cross Sections and Asymmetries for Longitudinally and Transversely Polarized Leptons and Hadrons in Deep Inelastic Scattering}
\author{Paul Anderson}
\affiliation{William and Mary, Williamsburg, VA 23185, USA}
\author{Douglas Higinbotham}
\affiliation{Thomas Jefferson National Accelerator Facility, Newport News, VA 23606, USA}
\author{Sonny Mantry}
\affiliation{University of North Georgia, Dahlonega, GA 30597, USA}
\author{Xiaochao Zheng}
\affiliation{University of Virginia, Charlottesville, VA 22904, USA}

\date{\today}

\begin{abstract}
    Studies of the Deep Inelastic Scattering (DIS) have provided fundamental information of the nucleon structure for decades. 
    The electron-ion collider (EIC) will be the first collider capable of DIS study with both polarized lepton and polarized hadron beams, providing the possibility of accessing new electroweak structure functions of the nucleon. 
    In this work, we completed the DIS cross section derivations for both longitudinally and transversely polarized leptons and hadrons, with no approximations made, and with all three contributions – $\gamma \gamma, \gamma Z, ZZ$ – included.  These results were derived using primarily tensor algebra and Feynman calculus, starting from previously established leptonic and hadronic tensors and carry out their contraction. 
    Our results are presented in terms of both spin-averaged and spin-dependent cross sections, allowing direct comparison with experimentally measured cross sections and their asymmetries. 
    We include also in our discussion comparisons of different conventions that exist in the literature.

    \begin{center}
         Presented at DIS2023: XXX International Workshop on Deep-Inelastic Scattering and Related Subjects, Michigan State University, USA, 27-31 March 2023
    \end{center}
\end{abstract}

\maketitle

\section{Introduction}

\begin{figure}[ht]
    \centering
    \includegraphics[scale = 0.10]{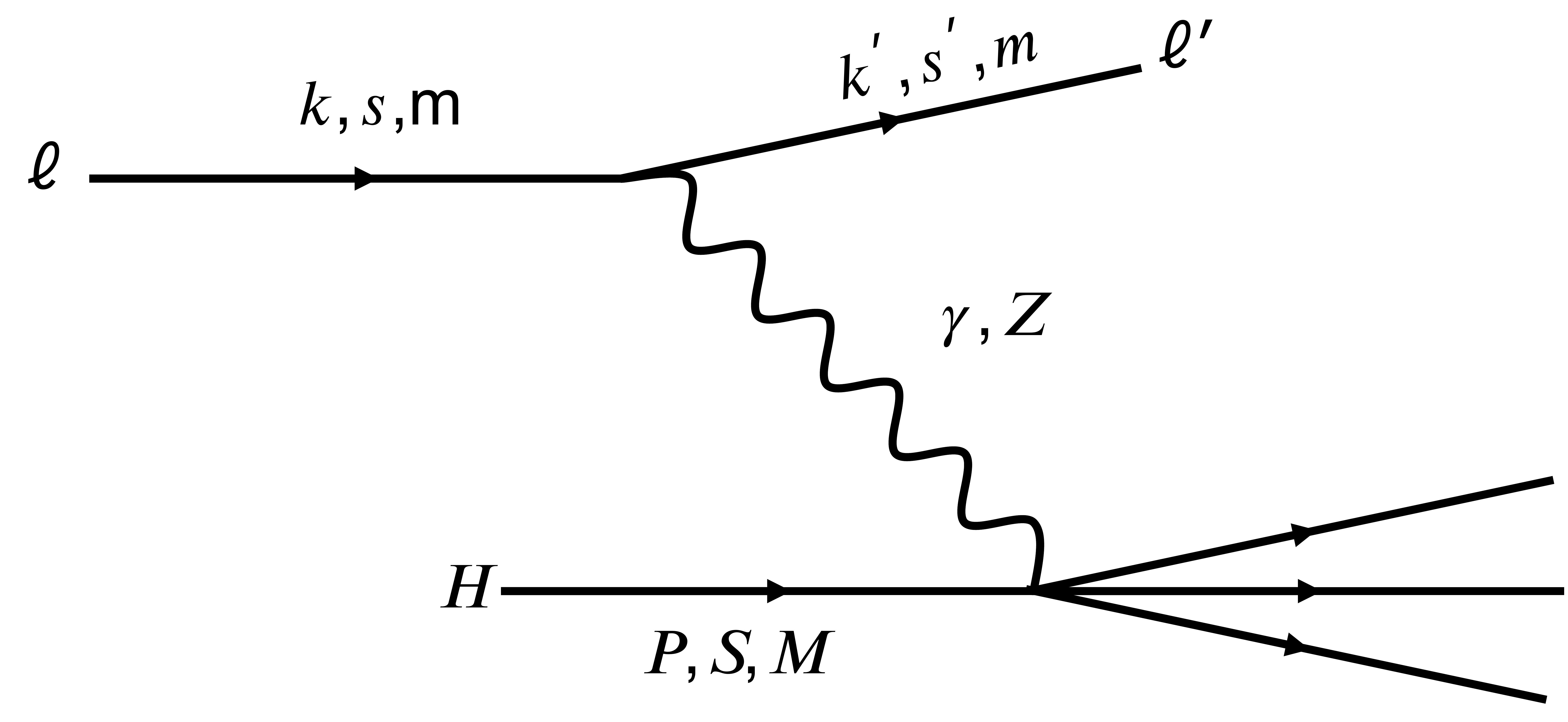}
    \caption{Feynman diagram for lepton-hadron collision.
    }
    \label{fig:FeynmanDiagram}
\end{figure}

With the construction of the Electron-Ion Collider (EIC) underway, it will soon be possible to perform Deep Inelastic Scattering (DIS) studies at a collider setting where both lepton and hadron beams have a spin polarization. Previous collider facilities have only had the capability to polarize at most one of the particles involved in the collision, while the kinematic reach of fixed-target facilities was limited. As such, there was previously no need for a complete and explicit cross section formalism with full spin degree of freedom and that includes all  electroweak terms. With the EIC, this has changed. In order to derive the cross sections and asymmetries for lepton-hadron collisions in the DIS regime where both particles have a spin polarization, we start with defining all 4-momenta and spin vectors as shown in Fig.~\ref{fig:FeynmanDiagram}. Here, $m (M)$ is the mass of the lepton (hadron), $E (E')$ and $k (k')$ are the initial (final) energy and momentum of the lepton, respectively, $s (s')$ is the initial (final) spin of the lepton, $S$ is the initial spin and $P$ the initial momentum of the hadron. We will use the commonly used notations: $q := k - k'$ is the momentum transfer; $Q^2 := - q^2$ is the negative square momentum transfer; $x := \frac{Q^2}{2(P \cdot q)}$ is the Bjorken scaling variable; and $y := \frac{P \cdot q}{P \cdot k}$ is the inelasticity. We also have the normalization $s^2 = S^2 = -1$ for the lepton and the hadron spins, respectively.

We focus on the following three specific spin polarization cases and provide the results.  
\begin{table}[ht]
    \centering
    \begin{tabular}{|c|c|}
        \hline
        Lepton Polarization & Hadron Polarization \\
        \hline
        \multirow{2}{*}{Longitudinal} & Longitudinal  \\
        \cline{2 - 2}
        & Transverse  \\
        \hline{Transverse} & Unpolarized  \\
        \hline
    \end{tabular}
    \caption{List of spin cases presented in this work. }
    \label{tab:spincases}
\end{table}
Among these spin cases, DIS between longitudinally polarized lepton and longitudinally or transversely polarized hadrons is the most relevant for JLab and EIC. DIS between transversely polarized leptons and unpolarized hadron is included because this is one of the primary background contribution for precision parity-violating electron scattering (PVES) experiments. Additionally, we have also derived the general case with arbitrary spin polarization direction of both the lepton and hadron, but will not include the results here due to space limitation. 

\subsection{Longitudinally Polarized Lepton}

\subsubsection{Longitudinally Polarized Hadron}

For longitudinally polarized leptons, we start with the leptonic tensor given in Eq.~(2.2.2) of~\cite{Anselmino:1994gn} (Anselmino et al., 1995),
\begin{equation}
    L_{\mu \nu}^{(\gamma)} = 2 \left[ k_\mu k_\nu' + k_\mu' k_\nu - (k \cdot k') g_{\mu \nu} - i \lambda_\ell \varepsilon_{\mu \nu \alpha \beta} k^\alpha k'^\beta \right]~,\label{eq:tensorL_longL}
\end{equation}
with Eqs.~(2.2.2-4) of~\cite{Anselmino:1994gn} re-written as: 
\begin{equation}
    L_{\mu \nu}^{(j)} = C^j L_{\mu \nu}^\gamma~,\\
\end{equation}
where $j = \gamma, \gamma Z$, and $Z$. The relevant conventions include: $(1, -1, -1, -1)$ for the metric tensor $g_{\mu \nu}$, $\varepsilon_{0 1 2 3} = 1 = -\varepsilon^{0 1 2 3}$ for the Levi-Civita symbol, and $\lambda_\ell = \pm 1$ for the left- and right-handed lepton helicities, which differs from the convention $\lambda_\ell = \pm \frac{1}{2}$ of~\cite{Anselmino:1994gn}. It should be noted that the tensor of Eq.~(\ref{eq:tensorL_longL}) is based on the assumption of a fast-moving lepton, $s^\mu \approx \frac{ \lambda_\ell}{m} k^\mu$ and a negligible lepton mass.
For the $C$ coefficients,  $C^\gamma = 1$, 
\begin{equation}\label{eqn:Cs}
C^{\gamma Z} = -(g_V - \lambda_\ell g_A),\textrm{~and~} C^Z = (g_V - \lambda_\ell g_A)^2~, 
\end{equation}
where $g_V=-\frac{1}{2}+2\sin^2\theta_W$ with $\theta_W$ the weak mixing angle and $g_A=-\frac{1}{2}$ are the vector and axial-vector couplings of the lepton to the $Z$ boson, respectively. Note that $C^{\gamma Z}$ differs by a minus sign from~\cite{Anselmino:1994gn}, arising from the charge of the electron being left out in~\cite{Anselmino:1994gn}.

For the hadronic tensor, we start from Eq.~(2.2.7) of~\cite{Anselmino:1994gn}, but with the sign of $g_4$ term flipped~\cite{Martini:2012thesis} to be consistent with the parton model interpretation, $g_4^j=2xg_5^j$, and Eq.(18.18) of Particle Data Group~\cite{PDG}. 

\vspace*{-0.6cm}
\begin{equation}\label{eqn:hadronictensor}
    \begin{split}
        &\frac{1}{2 M} W^{\mu \nu (j)} 
        = - \frac{g^{\mu \nu}}{M} F_1^j + \frac{P^\mu P^\nu}{M (P \cdot q)} F_2^j \\
        &+ i \frac{\varepsilon^{\mu \nu \alpha \beta}}{2 (P \cdot q)} \left( \frac{P_\alpha q_\beta}{M} F_3^j + 2 q_\alpha S_\beta g_1^j - 4 x P_\alpha S_\beta g_2^j \right) \\
        &- \frac{P^\mu S^\nu + S^\mu P^\nu}{2 (P \cdot q)} g_3^j - \frac{S \cdot q}{(P \cdot q)^2} P^\mu P^\nu g_4^j + \frac{S \cdot q}{P \cdot q} g^{\mu \nu} g_5^j~.
    \end{split}
\end{equation}
Note that the hadronic tensor of Eq.~(18.6) in~\cite{PDG} is $1/2$ of Eq.~(\ref{eqn:hadronictensor}), and the sign for both $g_{4,5}$ are the opposite. The $F_{1,2,3}$ and $g_{1,4,5}$ can be interpreted in the parton model~\cite{PDG}. One could in addition set $F_3^\gamma = g_3^\gamma = g_4^\gamma = g_5^\gamma = 0$ based on the current understanding that parity is conserved in electromagnetic interactions.

The neutral-current (NC) cross sections are formed from leptonic and hadronic tensors as: 
\begin{equation}\label{eqn:generalcrossseciton}
    \frac{\dd^2 \sigma_{\textrm{NC}}}{dx dy} = \frac{\pi y \alpha^2}{Q^4} \sum_{j = \gamma, \gamma Z, Z} \eta^j L_{\mu \nu}^{(j)} W^{\mu \nu (j)} 
\end{equation}
where $\eta^\gamma = 1$,
\begin{equation}
     \eta^{\gamma Z} = \left( \frac{G M_Z^2}{2 \sqrt{2} \pi \alpha} \right) \left( \frac{Q^2}{Q^2 + M_Z^2} \right); \mathrm{~and~} \eta^Z = \left( \eta^{\gamma Z} \right)^2~.
\end{equation}
Here, $\alpha$ is the fine structure constant, $M_Z$ is the mass of the $Z^0$ boson, and $G$ is the Fermi coupling constant. The factor $\frac{\pi y\alpha^2}{Q^4}$ of Eq.~(\ref{eqn:generalcrossseciton}) differs from PDG by factor two due to the different definitions of the hadronic tensor, as explained above. 

Our next step was to carry out tensor contraction for longitudinally polarized hadrons. The results involve a number of 4-vector dot products, most of which commonly used in DIS. However, a few complicated terms appear that are worth listing: $\frac{(1, - \hat{P}) \cdot k}{E + |\Vec{P}|}    \approx 0$, $k \cdot S = \frac{\lambda_H (P \cdot k)}{M}$, $k' \cdot S = \frac{\lambda_H (P \cdot k)}{M} \left( 1 - y - \frac{2 M^2 x^2 y^2}{Q^2} \right)$, $q \cdot S = \frac{\lambda_H (P \cdot k)}{M} \left( 1 + \frac{2 M^2 x^2 y}{Q^2} \right)$, where $\lambda_H=\pm 1$ is the helicity of the hadron and the lepton mass has been neglected. Then tensor contraction is: 
\begin{eqnarray}
       &&\hspace*{-0.4cm} L_{\mu \nu}^{(\gamma)} W^{\mu \nu (j)} \nonumber\\
       &=& 4 Q^2 F_1^j + \frac{4 Q^2}{x y^2} \left(1 - y - \frac{M^2 x^2 y^2}{Q^2} \right) F_2^j\nonumber\\
       && - \frac{2 \lambda_\ell Q^2 (2- y)}{y} F_3^j \nonumber\\
        &&+ \frac{4 \lambda_\ell \lambda_H Q^2}{y} \left( 2 - y - \frac{2 M^2 x^2 y^2}{Q^2} \right) g_1^j - 32 M^2 x^2 \lambda_\ell \lambda_H g_2^j \nonumber\\
        &&- \frac{4 Q^2 \lambda_H}{x y^2} \left( 1 - y - \frac{M^2 x^2 y^2}{Q^2} \right) g_3^j \nonumber\\
        &&- \frac{4 Q^2 \lambda_H}{x y^2} \left( 1 + \frac{2 M^2 x^2 y}{Q^2} \right) \left( 1 - y - \frac{M^2 x^2 y^2}{Q^2} \right) g_4^j \nonumber\\
        &&- 4 \lambda_H Q^2 \left( 1 + \frac{2 M^2 x^2 y}{Q^2} \right) g_5^j~.\label{eqn:longelonghtensorcontraction}
\end{eqnarray}
For $L^{\gamma Z,Z}W^j$, we multiply the RHS of Eq.~(\ref{eqn:longelonghtensorcontraction}) by the $C^j$. Finally, we note that coefficients for the $g_3$ term differ from~\cite{PDG}.

A cross section decomposition based on its spin dependence can be performed in a similar manner to that of~\cite{Boughezal:2022pmb}. Abbreviating $\frac{\dd^2\sigma}{\dd x \dd y}$ as $\dd\sigma$: 
\begin{equation}
    \begin{split}\label{eqn:SMEFTasymmetry}
        \dd\sigma_0 &= \frac{1}{4} \Big( \dd\sigma |_{++} + \dd\sigma |_{+-} + \dd\sigma |_{-+} + \dd\sigma |_{--} \Big)~, \\
        \dd\sigma_\ell &= \frac{1}{4} \Big( \dd\sigma |_{++} + \dd\sigma |_{+-} - \dd\sigma |_{-+} - \dd\sigma |_{--} \Big)~,\\
        \dd\sigma_H &= \frac{1}{4} \Big( \dd\sigma |_{++} - \dd\sigma |_{+-} + \dd\sigma |_{-+} - \dd\sigma |_{--} \Big)~, \\
        \dd\sigma_{\ell H} &= \frac{1}{4} \Big( \dd\sigma |_{++} - \dd\sigma |_{+-} - \dd\sigma |_{-+} + \dd\sigma |_{--} \Big)~.
    \end{split}
\end{equation}
where the first $\pm$ refer to the lepton helicity $\lambda=\pm 1$ and the second to the hadron helicity $\lambda_H=\pm 1$.
 
We present in the following results from explicit calculation, with $\Upsilon = 1 - y - \frac{M^2 x^2 y^2}{Q^2}, \zeta = 2 - y - \frac{2 M^2 x^2 y^2}{Q^2}, \xi = 1 - y - \frac{M^2 x^2 y}{Q^2}$, and $\Xi = 1 + \frac{2 M^2 x^2 y}{Q^2}$: 
\begin{eqnarray}
        \dd\sigma_0 &=& \frac{4 \pi y \alpha^2}{Q^2} \left[ F_1^\gamma - \eta^{\gamma Z} g_V F_1^{\gamma Z} + \eta^Z (g_V^2 + g_A^2) F_1^Z \right] \nonumber\\
        &+& \frac{4 \pi \alpha^2}{x y Q^2} \Upsilon \left[ F_2^\gamma - \eta^{\gamma Z} g_V F_2^{\gamma Z}  + \eta^Z (g_V^2 + g_A^2) F_2^Z \right] \nonumber\\
        &-& \frac{2 \pi \alpha^2 (2 - y)}{Q^2} \left( \eta^{\gamma Z} g_A F^{\gamma Z}_3 - 2 \eta^Z g_V g_A F^Z_3 \right)~,
    \label{eqn:longelonghasymmetry0}
\end{eqnarray}
\begin{eqnarray}
    \dd\sigma_\ell &=& \frac{4 \pi y \alpha^2}{Q^2} \left( \eta^{\gamma Z} g_A F^{\gamma Z}_1 - 2 \eta^Z g_V g_A F^Z_1 \right) \nonumber\\
    &+& \frac{4 \pi \alpha^2}{x y Q^2} \Upsilon \left( \eta^{\gamma Z} g_A F^{\gamma Z}_2 - 2 \eta^Z g_V g_A F^Z_2\right) \label{eqn:longelonghasymmetrye}\\
    &-& \frac{2 \pi \alpha^2 (2 - y)}{Q^2} \left[ - \eta^{\gamma Z} g_V F_3^{\gamma Z} + \eta^Z (g_V^2 + g_A^2) F_3^Z \right]~,\nonumber
\end{eqnarray}
\begin{eqnarray}
    \dd\sigma_H &=& \frac{4 \pi \alpha^2}{Q^2} \zeta \left( \eta^{\gamma Z} g_A g^{\gamma Z}_1 - 2 \eta^Z g_V g_A g^Z_1 \right) \nonumber\\
    &-& \frac{16 \pi M^2 \alpha^2 x^2 y}{Q^4} \left( \eta^{\gamma Z} g_A g^{\gamma Z}_2 - 2 \eta^Z g_V g_A g^Z_2 \right) \nonumber\\
    &-& \frac{4 \pi \alpha^2}{x y Q^2} \Upsilon \left[- \eta^{\gamma Z} g_V g_3^{\gamma Z} + \eta^Z (g_V^2 + g_A^2) g_3^Z\right] \label{eqn:longelonghasymmetryh}\\
    &-& \frac{4 \pi \alpha^2}{x y Q^2} \Xi \Upsilon \left[- \eta^{\gamma Z} g_V g_4^{\gamma Z} + \eta^Z (g_V^2 + g_A^2) g_4^Z\right] \nonumber\\
    &-& \frac{4 \pi y \alpha^2}{Q^2} \Xi \left[- \eta^{\gamma Z} g_V g_5^{\gamma Z} + \eta^Z (g_V^2 + g_A^2) g_5^Z\right]~,\nonumber
\end{eqnarray}
\begin{eqnarray}
    \dd\sigma_{\ell H} &=& \frac{4 \pi \alpha^2}{Q^2} \zeta \left[g_1^\gamma - \eta^{\gamma Z} g_V g_1^{\gamma Z} + \eta^Z (g_V^2 + g_A^2) g_1^Z\right] \nonumber\\
    &-& \frac{16 \pi M^2 \alpha^2 x^2 y}{Q^4} \left[g_2^\gamma - \eta^{\gamma Z} g_V g_2^{\gamma Z} + \eta^Z (g_V^2 + g_A^2) g_2^Z\right] \nonumber\\
    &-& \frac{4 \pi \alpha^2}{x y Q^2} \Upsilon \left( \eta^{\gamma Z} g_A g^{\gamma Z}_3 - 2 \eta^Z g_V g_A g^Z_3 \right) \label{eqn:longelonghasymmetryeh}\\
    &-& \frac{4 \pi \alpha^2}{x y Q^2} \Xi \Upsilon \left( \eta^{\gamma Z} g_A g^{\gamma Z}_4 - 2 \eta^Z g_V g_A g^Z_4 \right) \nonumber\\
    &-& \frac{4 \pi y \alpha^2}{Q^2} \Xi \left( \eta^{\gamma Z} g_A g^{\gamma Z}_5 - 2 \eta^Z g_V g_A g^Z_5 \right)~.\nonumber
\end{eqnarray}

\subsubsection{Transversely Polarized Hadron}

The case of longitudinally polarized lepton and transversely polarized hadron was considered next. The calculation is almost the same as the previous section, except the hadron spin $S$ changed, along with all dot products involving $S$. The tensor contraction was found to be:
\begin{eqnarray}
    L_{\mu \nu}^{(\gamma)} W^{\mu \nu (j)} &=& 4 Q^2 F_1^j + \frac{4 Q^2}{x y^2} \Upsilon F_2^j -\frac{4 \lambda_\ell Q^2 (2 - y)}{y} F_3^j\nonumber\\
    &+& 16 M \lambda_\ell x (k' \cdot S) g_1^j + \frac{32 M x \lambda_\ell}{y} (k' \cdot S) g_2^j\nonumber\\
    &-& \frac{4 M}{y} (k' \cdot S) g_3^j \nonumber\\
    &+& \frac{8 M}{y^2} \Upsilon (k' \cdot S) g_4^j + 8 M x (k' \cdot S) g_5^j~.
\end{eqnarray}
Note that for $g_4$ and $g_5$, the dot product $q \cdot S = (k \cdot S) - (k' \cdot S)$ was needed. We first note that $S = (0, \cos(\varphi), \sin(\varphi), 0)$ -- in the convention where the first number is the time component -- in the rest frame of the hadron, and is invariant under Lorentz boosts in the $z$-direction. 
One straightforward way to calculate the $q\cdot S$ product is thus to use the rest frame of the lepton where $k = (E, \Vec{0})$. We found $k \cdot S = 0$ and $q \cdot S = - k' \cdot S$.\bigskip

When considering the $k' \cdot S$ term, two possible bases could be considered. The first is to let $S$ along an arbitrary direction in the $x,y$-plane, while $k'$ only possesses components in the $x, z$ plane. The second is to let $S$ be fixed along the $x$-axis, while $k'$ is entirely arbitrary. In short, the decision is whether to have an azimuthal angle $\varphi$ on $S$ or $k'$. The vectors for the first case are: $k' = \left( E', \big| \Vec{k}' \big| \sin(\theta), 0, \big| \Vec{k}' \big| \cos(\theta) \right)$ and $S = \left(0, \cos(\varphi), \sin(\varphi), 0\right)$, 
while for the second case: $k' = \left(E', \big| \Vec{k}' \big| \cos(\varphi) \sin(\theta), \big| \Vec{k}' \big| \sin(\varphi) \sin(\theta), \big| \Vec{k}' \big| \cos(\theta) \right)$ and $S = (0, 1, 0, 0)$~.

\begin{figure}[!ht]
    \includegraphics[width=0.3\textwidth]{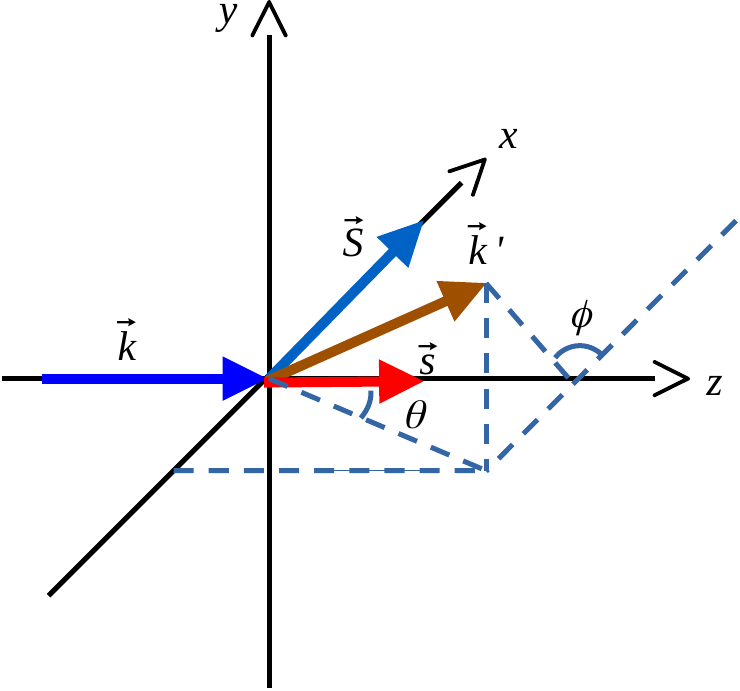}
    \caption{The convention chosen for longitudinally polarized lepton and transversely polarized hadron. The $x$ axis is chosen to be along the hadron spin $\vec S$, while $\vec k'$ is attributed a polar angle $\theta$ and an azimuthal angle $\phi$. }
    \label{fig:longetransh}
\end{figure}

The second basis was chosen for this work, see Fig.~\ref{fig:longetransh} for the definition of spin orientations and angles. The cross sections of  Eq.~(\ref{eqn:SMEFTasymmetry}) were then generalized, with $\lambda_H = \pm 1$ replaced with $\pm S$:
\vspace*{-0.6cm}
\begin{eqnarray}
   \dd\sigma_0 &=& \frac{1}{4} \Big( \dd\sigma |_{+, S} + \dd\sigma |_{+, -S} + \dd\sigma |_{-, S} + \dd\sigma |_{-, -S} \Big) \nonumber\\
   \dd\sigma_\ell &=& \frac{1}{4} \Big( \dd\sigma |_{+, S} + \dd\sigma |_{+, -S} - \dd\sigma |_{-, S} - \dd\sigma |_{ -, -S} \Big) \nonumber\\
   \dd\sigma_H &=& \frac{1}{4} \Big( \dd\sigma |_{+, S} - \dd\sigma |_{+, -S} + \dd\sigma |_{-, S} - \dd\sigma |_{ -, -S} \Big) \nonumber\\
   \dd\sigma_{\ell H} &=& \frac{1}{4} \Big( \dd\sigma |_{ +, S} - \dd\sigma |_{+, -S} - \dd\sigma |_{ -, S} + \dd\sigma |_{-, -S} \Big)\nonumber
\end{eqnarray}
The results are:
\vspace*{-0.1cm}
\begin{eqnarray}
    \dd\sigma_0 &=& \frac{4 \pi y \alpha^2}{Q^2} \left[ F_1^\gamma - \eta^{\gamma Z} g_V F_1^{\gamma Z} + \eta^Z (g_V^2 + g_A^2) F_1^Z \right] \nonumber \\
    &+& \frac{4 \pi \alpha^2}{x y Q^2} \Upsilon \left[ F_2^\gamma - \eta^{\gamma Z} g_V F_2^{\gamma Z} + \eta^Z (g_V^2 + g_A^2) F_2^Z \right] \nonumber \\
    &-& \frac{2 \pi \alpha^2 (2 - y)}{Q^2} \left( \eta^{\gamma Z} g_A F^{\gamma Z}_3 - 2 \eta^Z g_V g_A F^Z_3 \right)~,\label{eqn:longetranshasymmetry0}
\end{eqnarray}
\vspace*{-0.2cm}
\begin{eqnarray}
    \dd\sigma_\ell &=& \frac{4 \pi y \alpha^2}{Q^2} \left( \eta^{\gamma Z} g_A F^{\gamma Z}_1 - 2 \eta^Z g_V g_A F^Z_1 \right) \nonumber\\
    &+& \frac{4 \pi \alpha^2}{x y Q^2} \Upsilon \left( \eta^{\gamma Z} g_A F^{\gamma Z}_2 - 2 \eta^Z g_V g_A F^Z_2\right) \label{eqn:longetranshasymmetrye}\\
    &-& \frac{2 \pi \alpha^2 (2 - y)}{Q^2} \left[ - \eta^{\gamma Z} g_V F_3^{\gamma Z} + \eta^Z (g_V^2 + g_A^2) F_3^Z \right]~,\nonumber
\end{eqnarray}
\vspace*{-0.2cm}
\begin{eqnarray}
   \dd\sigma_H &=& \frac{8 \pi \alpha^2 M x y k_1'}{Q^4} \left( \eta^{\gamma Z} g_A g^{\gamma Z}_1 - 2 \eta^Z g_V g_A g^Z_1 \right)\nonumber\\
   &+& \frac{16 \pi \alpha^2 M x k_1'}{Q^4} \left( \eta^{\gamma Z} g_A g^{\gamma Z}_2 - 2 \eta^Z g_V g_A g^Z_2 \right) \nonumber\\
   &-& \frac{4 \pi \alpha^2 M k_1'}{Q^4} \left[- \eta^{\gamma Z} g_V g_3^{\gamma Z} + \eta^Z (g_V^2 + g_A^2) g_3^Z\right] \label{eqn:longetranshasymmetryh}\\
   &+& \frac{8 \pi \alpha^2 M k_1'}{y Q^4} \Upsilon \left[- \eta^{\gamma Z} g_V g_4^{\gamma Z} + \eta^Z (g_V^2 + g_A^2) g_4^Z\right] \nonumber\\
   &+& \frac{8 \pi M x y \alpha^2 k_1'}{Q^4} \left[- \eta^{\gamma Z} g_V g_5^{\gamma Z} + \eta^Z (g_V^2 + g_A^2) g_5^Z\right]~,\nonumber
\end{eqnarray}
and
\begin{eqnarray}
    \dd\sigma_{\ell H} &= \frac{8 \pi \alpha^2 M x y k_1'}{Q^4} \left[g_1^\gamma - \eta^{\gamma Z} g_V g_1^{\gamma Z} + \eta^Z (g_V^2 + g_A^2) g_1^Z\right] \nonumber\\
    &+ \frac{16 \pi \alpha^2 M x k_1'}{Q^4} \left[g_2^\gamma - \eta^{\gamma Z} g_V g_2^{\gamma Z} + \eta^Z (g_V^2 + g_A^2) g_2^Z\right] \nonumber\\
    &- \frac{4 \pi \alpha^2 M k_1'}{Q^4} \left( \eta^{\gamma Z} g_A g^{\gamma Z}_3 - 2 \eta^Z g_V g_A g^Z_3 \right) \label{eqn:longetranshasymmetryeh}\\
    &+ \frac{8 \pi \alpha^2 M k_1'}{y Q^4} \xi \left( \eta^{\gamma Z} g_A g^{\gamma Z}_4 - 2 \eta^Z g_V g_A g^Z_4 \right) \nonumber\\
    &+ \frac{8 \pi M x y \alpha^2 k_1'}{Q^4} \left( \eta^{\gamma Z} g_A g^{\gamma Z}_5 - 2 \eta^Z g_V g_A g^Z_5 \right)~,\nonumber
\end{eqnarray}
where $k_1' := k' \cdot S
= |\Vec{k}'| \cos(\varphi) \sin(\theta)$.

\subsection{Transversely Polarize Lepton}
The case of transversely polarized lepton was discussed briefly in~\cite{Anselmino:1994gn} and it was stated that the lepton spin-dependent cross section difference is proportional to the lepton mass and thus is negligible. We derive below explicitly the result for DIS of transversely polarized lepton and unpolarized hadron case because this could potentially be a background for parity-violating electron scattering (PVES) experiments, which are now carried out to higher and higher precision at a number of fixed-target facilities. 

\subsubsection{Unpolarized Hadron Case}

\begin{figure}[!b]
\includegraphics[width=0.3\textwidth]{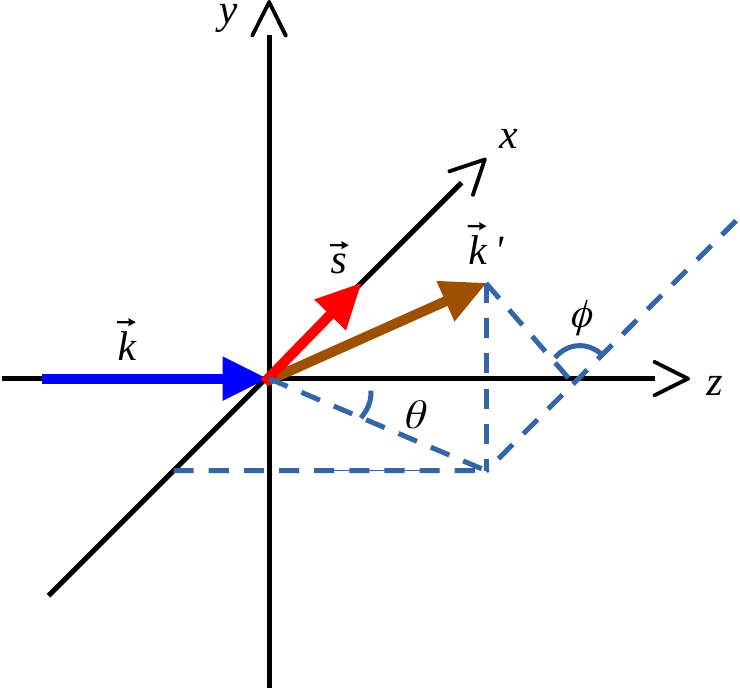}
    \caption{The convention chosen for transversely polarized
lepton scattering. The $x$ axis is chosen to be along the lepton spin $\vec s$, while $\vec k'$ is attributed a polar angle $\theta$ and an azimuthal angle $\phi$. The hadron spin $\vec S$ is averaged in the calculation and is not shown.}
    \label{fig:transenoh}
\end{figure}
For the transversely polarized lepton case, the assumption of the fast-moving electron was dropped, and the lepton mass was kept. The leptonic tensor for transversely polarized leptons is:
\begin{eqnarray}
    && L_{\mu \nu}^{(\gamma)} = 2(k_\mu k'_\nu + k_\nu k'_\mu) \nonumber \\
    && \hspace*{1cm} + 2 g_{\mu \nu}(m^2 - k \cdot k') \nonumber \\
    && \hspace*{1cm} + 2 m s^\alpha (k - k')^\beta \varepsilon_{\mu \nu \alpha \beta} i \\
    && L_{\mu \nu}^{(\gamma Z)} = - 2 g_V L_{\mu \nu}^{(\gamma)} \nonumber \\
    && \hspace*{1cm} - 4 g_A \Big[m (k' \cdot s) g_{\mu \nu} - m (k_\mu' s_\nu + k_\nu' s_\mu) \nonumber \\
    && \hspace*{1cm} + k^\alpha k'^\beta \varepsilon_{\mu \nu \alpha \beta} i\Big] \\
    && L_{\mu \nu}^{(Z)} = (g_V^2 + g_A^2) L_{\mu \nu}^{(\gamma)} - 4 g_A^2 m^2 g_{\nu \nu} \nonumber \\
    && \hspace*{1cm} - 4 g_A^2 m s^\alpha k^\beta \varepsilon_{\mu \nu \alpha \beta} i + 4 g_V g_A \Big[ m(k' \cdot s) g_{\mu \nu} \nonumber \\
    && \hspace*{1cm} - m(k'_\mu s_\nu + k'_\nu s_\mu) + k^\alpha k'^\beta \varepsilon_{\mu \nu \alpha \beta} i \Big]~.
\end{eqnarray}
Note that unlike longitudinally polarized lepton case, the $\gamma Z$ and $Z$ components of the leptonic tensor can no longer be written as proportional to the $\gamma$ component. 

To obtain the case for unpolarized hadrons, the average was taken over the hadron spin $S$ and $-S$, and we form only two cross section combinations: 
\begin{align}
    \dd\sigma_0 &= \frac{1}{2} \Big( \dd\sigma |_s + \dd\sigma |_{-s} \Big)~, \\
    \dd\sigma_\ell &= \frac{1}{2} \Big( \dd\sigma |_s - \dd\sigma |_{-s} \Big)~,
\end{align}
where $s$ is the lepton spin.

Similar to the longitudinally polarized lepton and transversely polarized hadron case, there were two potential options for the basis of $k'$ and $s$. If we choose to fix the lepton spin along $x$ and let $\vec k'$ direction be arbitrary, we can write $s = (0, 1, 0, 0)$ and $k' = \left( E', |\Vec{k}'| \cos(\varphi) \sin(\theta), |\Vec{k}'| \sin(\varphi) \sin(\theta), \Vec{k}' \cos(\theta) \right)$, see Fig.~\ref{fig:transenoh}. Since $s$ is invariant under Lorentz boosts in the $z$-direction, we can utilize that in the hadron rest frame $P = (M, \Vec{0})$ and conclude $P \cdot s = 0$. This specific result simplies the cross section results, which are found to be:
\begin{eqnarray}
    \dd\sigma_0 &&= - \frac{4 \pi y \alpha^2}{Q^4} \Big\{ \left(2 m^2 - Q^2 \right) F_1^\gamma \nonumber \\
    &&- 2 \left( 2 m^2 - Q^2 \right) g_V \eta^{\gamma Z} F_1^{\gamma Z} \nonumber \\
    &&+ \left[ \left( g_V^2 + g_A^2 \right) \left( 2 m^2 - Q^2 \right) - 8 g_A^2 m^2 \right] \eta^Z F_1^Z \Big\} \nonumber \\
    &&+ \frac{4 \pi \alpha^2}{x y Q^2} \Bigg\{ \Upsilon F_2^\gamma - 2 g_V \eta^{\gamma Z} \Upsilon F_2^{\gamma Z} \nonumber \\
    &&+ \eta^Z \left[ \left( g_V^2 + g_A^2 \right) \Upsilon - \frac{4 g_A^2 m^2 M^2 x^2 y^2}{Q^4} \right] F_2^Z \Bigg\} \nonumber \\
    &&+ \frac{4 \pi \alpha^2 g_A (y - 2)}{Q^2} \left( \eta^{\gamma Z} F_3^{\gamma Z} - g_V \eta^Z F_3^Z \right)
\end{eqnarray}
\begin{eqnarray}
    \dd\sigma_\ell &&= \frac{16 \pi y \alpha^2 m g_A (k' \cdot s)}{Q^4} \left( \eta^{\gamma Z} F_1^{\gamma Z} - g_V \eta^Z F_1^Z \right) \nonumber \\
    &&- \frac{16 \pi m M^2 x y \alpha^2 g_A (k' \cdot s)}{Q^6} \left( \eta^{\gamma Z} F_2^{\gamma Z} - g_V \eta^Z F_2^Z \right) \nonumber \\
    &&+ \frac{4 \pi \alpha^2 m (k' \cdot s)}{Q^4} \Big\{ - 2 y g_V \eta^{\gamma Z} F_3^{\gamma Z} \nonumber \\
    &&+ \left[ y \left( g_V^2 + g_A^2 \right) - 2 g_A^2 \right] \eta^Z F_3^Z \Big\}~.
\end{eqnarray}
Note that the lepton-spin dependent cross section difference $d\sigma_\ell$ is proportional to $k'\cdot s$ and thus would have a $\cos\phi$ modulation. 

In typical PVES experiments, the main parity-violating observable of interest is the cross section asymmetry between right- and left-handed incident electrons on an unpolarized target. In reality, the spin of the incident electron beam may have a transverse component, causing background signal due to two-photon exchange effects which are of pure electromagnetic nature. 
All PVES experiments thus had dedicated measurements of the beam ``normal'' single-spin asymmetry, $A_n$, to directly access this background using transversely polarized electron beam. In recent experiments with full azimuthal acceptance of the scattered electron, the $A_n$ is determined by fitting the $\sin\phi$ component of the observed cross section asymmetry associated with the electron beam spin flip, see e.g.~\cite{QWeak:2021jew,PREX:2021uwt}.  However, as can be seen above, there can be a $\cos\phi$ component caused by parity-violation.  Our results show that the parity-violating asymmetry of transversely polarized lepton is of the order of $\eta^{\gamma Z} m^2/Q^2\approx m^2/M_Z^2\approx 10^{-11}$, and is likely to be too small to be observed in such $A_n$ measurements.

\bigskip
\section{Discussion and Summary}

We have completed the DIS cross section formalism in the most complete form known to date, that includes lepton and hadron spin in arbitrary directions, all neutral current terms, and both lepton and hadron mass terms. 
These results should serve as a foundation for future asymmetry experiments at Jefferson Lab and the Electron Ion Collider. Results on the spin-dependent cross section asymmetry for transversely polarized electron case are also relevant for beam-normal single spin asymmetry measurements. From here, projections of the size of various asymmetries for possible experimental measurements at Jefferson Lab and EIC can be made, and feasibility of new measurements studied. 

\bigskip
\centerline{\bf{Code Availability}}
We made use of the FeynCalc package of Mathematica~\cite{MERTIG1991345, Shtabovenko_2016, Shtabovenko_2020} for cross-checking our calculations. The Mathematica code used can be downloaded from \url{https://www.dropbox.com/s/gwlkhibxerpf0jd/DIS%20Cross%20Sections.nb?dl=0}.

\bigskip
\acknowledgments{
    This work is supported in part by the U.S. Department of Energy, Office of Science, Office of Nuclear Physics under contract number DE–SC0014434 (U. of Virginia); and Office of Science, Office of Workforce Development for Teachers and Scientists (WDTS) under the Science Undergraduate Laboratory Internships Program. Travel support to the DIS2023 conference was provided by the organizers and the College of William \& Mary.
}

\bibliographystyle{JHEP}
\bibliography{dis_xs}

\end{document}